\documentclass[journal,transmag]{IEEEtran}

\usepackage[compress]{cite}
\usepackage{graphicx}
\usepackage{amssymb,amsmath,upgreek,url,lipsum,bm}
\usepackage[per-mode=symbol]{siunitx}
\usepackage{svg}
\usepackage{adjustbox}
\usepackage{flushend}
\usepackage{xcolor}
\usepackage{bm}

\newcommand{\ionspercm}{$\mathrm{ions/cm^2}$}
\newcommand{\electron}{e}

\begin{document}
\title{The Effect of Ga-Ion Irradiation on Sub-Micron-Wavelength\\ Spin Waves in Yttrium-Iron-Garnet Films}

% author name
\author{\IEEEauthorblockN{Johannes Greil\IEEEauthorrefmark{1}, Martina Kiechle\IEEEauthorrefmark{1}, Adam Papp\IEEEauthorrefmark{2}, Peter Neumann\IEEEauthorrefmark{3}, Zoltán Kovács\IEEEauthorrefmark{3}, Janos Volk\IEEEauthorrefmark{3},\\ Frank Schulz\IEEEauthorrefmark{4}, Sebastian Wintz\IEEEauthorrefmark{5}, Markus Weigand\IEEEauthorrefmark{5}, György Csaba\IEEEauthorrefmark{2}, Markus Becherer\IEEEauthorrefmark{1}}
\IEEEauthorblockA{\IEEEauthorrefmark{1}School of Computation, Information and Technology, Technical University of Munich, München, Germany}
\IEEEauthorblockA{\IEEEauthorrefmark{2}Faculty of Information Technology and Bionics, Pázmány Péter Catholic University, Budapest, Hungary}
\IEEEauthorblockA{\IEEEauthorrefmark{3} HUN-REN Center for Energy Research, Budapest, Hungary}
\IEEEauthorblockA{\IEEEauthorrefmark{4} Max-Plank-Institut für Intelligente Systeme, Stuttgart, Germany}
\IEEEauthorblockA{\IEEEauthorrefmark{5} Helmholtz-Zentrum Berlin (HZB), Berlin, Germany}
}

%Keep this as it is!!!
\IEEEtitleabstractindextext{%
\begin{abstract}
We investigate the effect of focused-ion-beam (FIB) irradiation on spin waves with sub-micron wavelengths in Yttrium-Iron-Garnet~(YIG) films. Time-resolved scanning transmission X-ray (TR-STXM) microscopy was used to image the spin waves in irradiated regions and deduce corresponding changes in the magnetic parameters of the film. We find that the changes of Ga$^+$ irradiation can be understood by assuming a few percent change in the effective magnetization $\bm{M_\mathrm{eff}}$ of the film due to a trade-off between changes in anisotropy and effective film thickness. Our results demonstrate that FIB irradiation can be used to locally alter the dispersion relation and the effective refractive index $\bm{n_\textrm{eff}}$ of the film, even for submicron wavelengths. To achieve the same change in $\bm{n_\textrm{eff}}$ for shorter wavelengths, a higher dose is required, but no significant deterioration of spin wave propagation length in the irradiated regions was observed, even at the highest applied doses.
%To achieve the same change in $\bm{n_\textrm{eff}}$ for shorter wavelengths, a higher dose is required, but we see no significant deterioration of the spin wave propagation length in the irradiated regions even at higher doses.

%We report on the achievable effective refractive-index changes and the corresponding dose levels and discuss the implications of our key findings. FIB irradiation was found to have a diminishing effect on the refractive index at shorter wavelengths, which is in agreement with our expectations based on the analysis of the dispersion curves of spin waves. Nevertheless, we observe long propagation lengths for comparably high doses, making sub-micron wavelength spin wave devices using FIB technology accessible. We discuss the challenges in downscaling optical elements for spin waves and provide a perspective on the anticipated limitations of the technology. %This report represents an important milestone towards miniaturization of spin-wave optics.
\end{abstract}
}

\maketitle

\section{Introduction}
Magnonics has seen explosive development in recent years, with many technological breakthroughs \cite{Chumak2022,zenbaa2024,Breitbach2023stimulated,Breitbach2024,Merbouche_2024True} and interesting device concepts \cite{wang2024allmagnonic,wang2024nanoscale,Casulleras2023generation,Flebus2024The}. 
As the field matures, it becomes gradually more important to develop industry-standard fabrication technologies that are versatile, flexible, scalable, and widely accessible. 
Since the focus is drifting toward applications, the complexity of the devices in a single magnonic unit (i.e., a device block without internal electrical conversion) is increasing, requiring more sophisticated patterning technologies. 
Recently, focused-ion-beam (FIB) irradiation of Yttrium-Iron-Garnet (YIG) has been demonstrated as a tool for fabricating optical elements for spin waves~\cite{Kiechle2023}. 
Instead of physically removing material to create patterns, this method introduces local changes in the crystal structure of YIG, effectively modifying the local dispersion properties of spin waves. 
In effect, gradient-index steering of spin waves becomes achievable for both classical optical geometries and inverse-designed scatterers~\cite{Kiechle2022,Kiechle2023}.
This technique has a high resolution, limited mainly by the vertical and lateral straggle of the ions, which is typically less than the film thickness or a few tens of nanometers. 
In research labs, it provides a rapid prototyping workflow, as it does not require lithography or a clean room environment. 
Moreover, it can be applied multiple times successively to further adjust the device characteristics. 
With regard to mass production, ion-irradiation can be adapted to production lines using  ion implanter technology combined with hard masks, which is a standard technology in modern CMOS production lines.

In this work, we were interested in finding out the trade-offs of using the technology of Ga$^+$-ion irradiation for spin waves with an order of magnitude smaller wavelengths, i.e., a few hundred nanometers instead of a few microns as demonstrated in \cite{Kiechle2023}.
For this purpose, we carried out time-resolved scanning transmission X-ray microscopy (TR-STXM) at the Maxymus end station~\cite{Weigand2022} of the BESSY II electron storage ring, operated by the Helmholtz-Zentrum Berlin für Materialien und Energie on samples with sub-micron stripline transducers to excite spin waves.
%For this, we applied for beamtime at the BESSY II synchrotron in Berlin for time-resolved soft X-ray imaging based on scanning transmission X-ray microscopy (STXM) at the MAXYMUS beamline \cite{Weigand2022} and fabricated samples with sub-micron stripline transducers to excite spin waves. 
%Unfortunately, most of our efforts were nulled by defects and bubbles at the YIG/GGG interface, which impeded the preparation of a membrane for X-ray transparency. 
%We measured wavelengths down to \SI{700}{\nano\meter}, unfortunately only for a single dose map. 
We measured wavelengths down to \SI{700}{\nano\meter} for a lateral map of several ion doses.
%Although the demonstration of spin-wave optical components was not successful for this reason, we could record the dose map. 
We obtained an assessment of the required doses for sub-micron wavelengths where even relatively high doses do not significantly deteriorate spin wave propagation lengths.  

\section{Methods}

\begin{figure}[b!]
    \centering
    \includegraphics[trim={0 0 2.5cm 0}, clip, width=0.95\columnwidth]{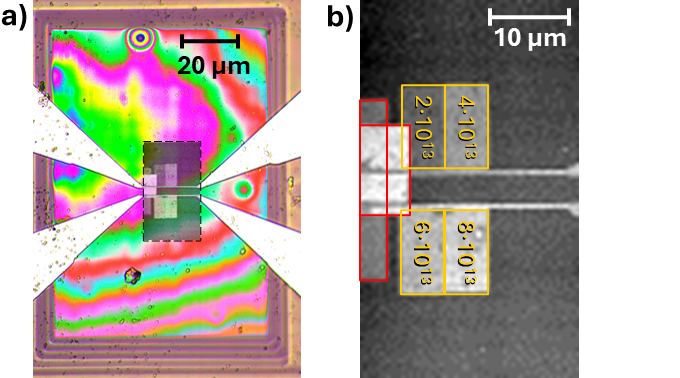}
    \caption{a) Optical microscope image of the sample geometry. The superimposed black-white image shows the image from b) for reference. b) Normalized STXM image of the investigated sample area. Regions irradiated by FIB are indicated with the corresponding dose in \ionspercm and a yellow frame. Regions indicated by a red frame were irradiated for alignment purposes, partially overlapping with the dose map.}
    \label{fig:geometry}
\end{figure}

In our experiment, we used a YIG film of \SI{100}{\nano\meter} thickness purchased from Innovent Jena.
The film was grown by means of liquid phase epitaxy on a GGG substrate~\cite{Dubs2020}, with an extracted intrinsic effective magnetization $M_\mathrm{eff}$ of \SI{133}{\kilo\ampere\per\meter}. 
%In our experiments, we used a \SI{100}{\nano\meter} thin YIG film with an extracted intrinsic effective magnetization M$_\mathrm{eff}$ of \SI{133}{\kilo\ampere\per\meter} on a GGG membrane.
For thinning the GGG substrate to a membrane thickness of several hundred nanometers in an area of \SI{135}{\micro\meter} by \SI{95}{\micro\meter}, mechanical polishing in combination with Ga$^+$ FIB milling, similar to the process described in \cite{Mayr2021} was used.
%The optical microscope image in Fig.~\ref{fig:geometry}a) shows the sample geometry with the supplied stripline transducer on the left and the unsupplied transducer on the right.
The optical microscope image in Fig.~\ref{fig:geometry}a) shows the sample geometry with the upper stripline transducer being supplied while the lower transducer is unsupplied.
%The stripline transducers were fabricated using a Raith\num{150} electron-beam lithography system and have a width of \SI{500}{\nano\meter}.
The stripline transducers with a width of \SI{500}{\nano\meter} were fabricated using electron-beam lithography, thermal evaporation of a Cr(\SI{8}{\nano\meter})/Cu(\SI{180}{\nano\meter})/Cr(\SI{8}{\nano\meter}) stack, and lift-off processing.
The backside-thinned YIG/GGG membrane shows thin-film interference for optical wavelengths, indicating a slight thickness variation of the overall membrane across the prepared window, which thus appears to be rainbow-colored. 
The brown nested rectangles around the membrane are thicker GGG remnants of the step-wise decreasing area in the backside etching process.
%The backside-thinned YIG/GGG membrane appears to be rainbow-colored, while the brown nested rectangles around the membrane are GGG remnants of the step-wise decreasing area in the backside etching process. 
%The brightness change in the intrinsic film areas in the XMCD contrast in Fig.~\ref{fig:geometry}b) and absolute STXM picture in Fig.~\ref{fig:snapshot}a) show that the thickness of the GGG membrane has a negligible variation in the evaluated region~\cite{Mayr2021}.
The brightness change in the absolute STXM picture in Fig.~\ref{fig:snapshot}a) shows that the thickness of the YIG/GGG membrane has a negligible variation~in the~evaluated~region~\cite{Mayr2021}.

For the Ga$^+$-ion irradiation of the YIG film, we used a \SI{50}{\kilo\volt} Micrion 9500 EX FIB tool.
The mean penetration depth of the accelerated Ga$^+$-ions at \SI{50}{\kilo\volt} is about \SI{25}{\nano\meter} according to SRIM simulations \cite{Ziegler2010SRIM} and the TEM measurements presented in the supplementals of~\cite{Kiechle2023}.
%The used dose levels in our experiment range from \SIrange{1e13}{8e13}{}\,\ionspercm{}, which is one order of magnitude higher than the experiments shown in \cite{Kiechle2022,Kiechle2023}.

%The experiments were conducted at the BESSY II synchrotron at the MAXYMUS  beamline using scanning transmission X-ray microscopy (STXM)~\cite{Weigand2022}. 
To achieve maximum X-ray magnetic circular dichroism (XMCD) contrast, the presented measurements were conducted at the central Fe $\mathrm{L_3}$ XMCD peak of YIG at a nominal photon energy of \SI{709.6}{\electron\volt}.
%The sample was mounted \SI{30}{\degree} tilted relative to the incident beam to apply an out-of-plane magnetic field from \SIrange{-250}{250}{\milli\tesla}~\cite{Nolle2012}.
The sample was mounted \SI{30}{\degree} tilted relative to the incident beam to gain sensitivity for forward volume (FV) spin waves, and magnetic fields were applied perpendicular to the sample plane in a range between \SIrange{-250}{250}{\milli\tesla}~\cite{Nolle2012}.

\section{Results}

\begin{figure}[tbh!]
    \centering
    \includegraphics[width=0.95\columnwidth]{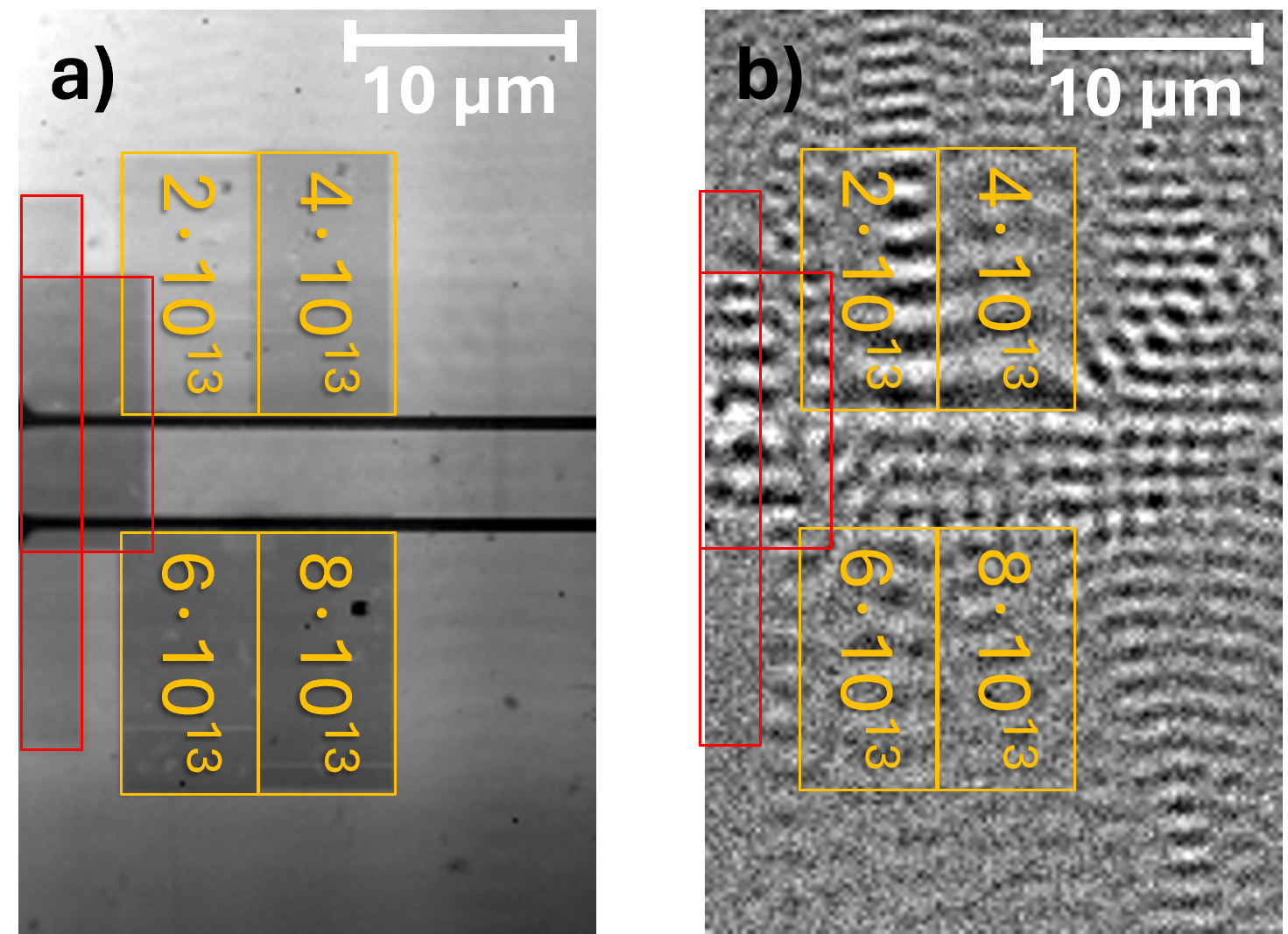}
    \caption{Recorded TR-STXM images of the spin-wave propagation for a bias field of $B = \SI{242.5}{\milli\tesla}$. A yellow frame indicates irradiated regions, and alignment regions with respect to the transducer are indicated by red frames. a) A recoded snapshot image before subtracting the static background. Irradiated regions are visible as darker areas in the static, partially magnetic contrast, with higher doses resulting in darker shades. b) Normalized TR-STXM snapshot, from the same experiment. The intensity in the lower boxes (higher doses) is typically smaller since the transducer does not excite waves directly in these regions. Intrinsic spin waves have to cover the distance to the unsupplied transducer and cross the boundary of the irradiated regions. These waves are also partially reflected from the boundary.}
    \label{fig:snapshot}
\end{figure}

We measured the wavelength of spin waves in the intrinsic YIG film and in irradiated regions at various bias fields in the FV configuration, i.e.,  with an out-of-plane bias field.
Exemplary snapshots of the absolute and normalized (divided by the time average) STXM measurement are shown in Fig.~\ref{fig:snapshot}a) and b) respectively, for a DC magnetic bias field of $B = \SI{242.5}{\milli\tesla}$.
Spin wave propagation can be detected in the four dose fields, whereas the lower two fields show less signal amplitude because the waves must pass the intrinsic region between the transducers.
Moreover, in Fig~\ref{fig:snapshot}b), the transition between the upper dose fields and the intrinsic film is visible, showing the abrupt change in wavelength when irradiating areas with sharp borders.

\begin{figure}[tbh!]
    \centering
    \includegraphics[width=0.9\columnwidth]{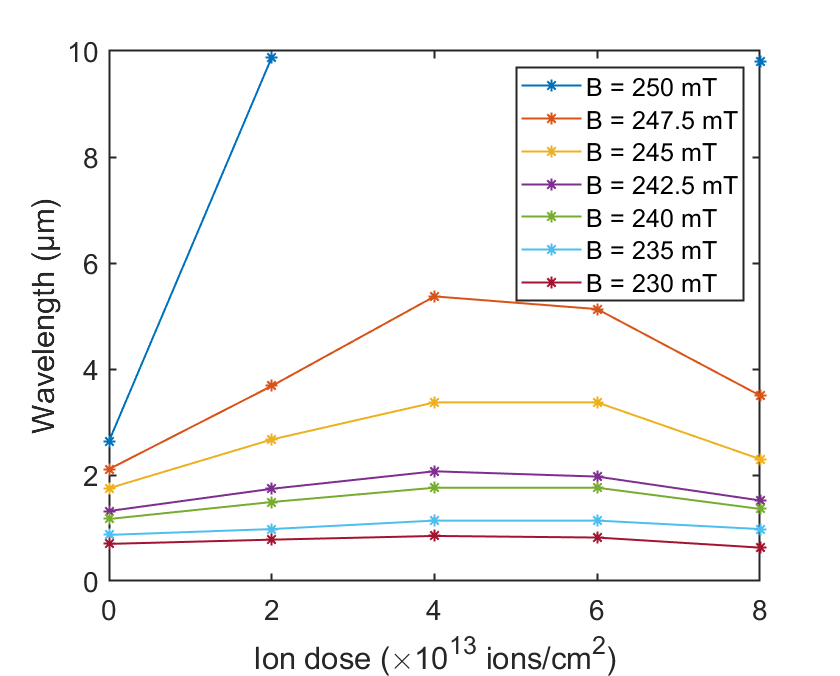}
    \caption{Measured spin wave wavelength vs. $\mathrm{Ga}^+$-ion dose. The colored curves correspond to different applied bias fields, as indicated in the legend. The wavelength initially increases in all cases as a result of the irradiation, and the maximum change in wavelength occurs around \SI{4e13}{}\,\ionspercm{}. For higher doses, the wavelength decreases again. Overall, the change in wavelength is smaller for shorter intrinsic wavelengths.}
    \label{fig:dosemap}
\end{figure}

The extracted wavelengths for the intrinsic and irradiated areas at different bias fields are plotted in Fig.~\ref{fig:dosemap}.
We observe a non-monotonic behavior as the wavelength first increases with the dose, but after a maximal change occurring between $4-6 \times 10^{13}$\,ions/cm$^2$, the trend reverses. 
This behavior was also observed in \cite{Kiechle2023}, but the wavelength in that case decreased initially and reached a local minimum. 
The two samples are similar in most parameters, except for the dose levels (one order of magnitude higher in the present study), the range of wavelengths excited (almost an order of magnitude lower in the present study), and the fact that the substrate was thinned down to a membrane for X-ray transparency.
The latter aspect could change the effect of strain in the system, which may occur at a comparatively large area across the membrane, while the Ga$^+$-FIB backside milling is not affecting the YIG layer~directly~\cite{Mayr2021}.

\begin{figure}[tbh!]
    \centering
    \includegraphics[width=0.95\columnwidth]{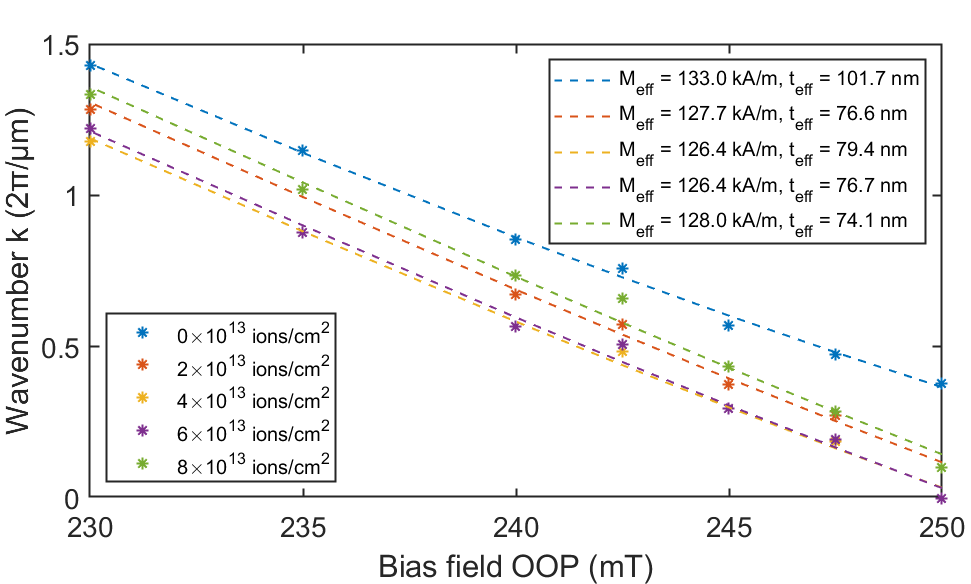}
    \caption{Measured dependency of the wavenumber on the bias field for the various applied dose levels. Dashed lines indicate analytical curves based on the Kalinikos and Slavin dispersion model with fitted effective magnetization and film thickness values indicated in the legend.}
    \label{fig:B_k}
\end{figure}

We have plotted the wavenumber for the different dose levels as a function of the bias field in Fig.~\ref{fig:B_k}. 
The field dependence is almost linear, which facilitates analyzing the data. 
We fitted analytical curves to the data using the dispersion formulas developed by Kalinikos and Slavin~\cite{Kalinikos1986}. 
Due to the almost symmetrical shape of the curves in Fig.~\ref{fig:dosemap}, the fitted wavenumber curves in Fig.~\ref{fig:B_k} appear in pairs (i.e., \SI{2e13}{}\,\ionspercm{} \& \SI{8e13}{}\,\ionspercm{} and \SI{4e13}{}\,\ionspercm{} \& \SI{6e13}{}\,\ionspercm{} result in almost the same extracted values, respectively).

To match the slope and offset of the experimental data lines, we fitted the effective magnetization and the effective magnetic thickness of the crystalline YIG film in the irradiated cases. 
The fitted thicknesses (approximately \SIrange{75}{80}{\nano\meter} instead of the original \SI{100}{\nano\meter}) are in line with the expected mean penetration depth of the Ga$^+$-ions. 
TEM measurements show that approximately the top \SI{25}{\nano\meter} of the crystal receives the most damage and gets amorphous~\cite{Kiechle2023}. The results suggest that in all the irradiated cases, the top \SI{25}{\nano\meter} is almost completely destroyed. 
Thus, the wavelength differences between the different dose levels are not due to a physical thickness change in the film.
In addition to the effective magnetic thickness change, the effective magnetization $M_\mathrm{eff}$ was found to decrease by approximately \SI{5}{\kilo\ampere\per\meter} and \SI{7}{\kilo\ampere\per\meter} for the irradiated regions, as also indicated in the legend of Fig.~\ref{fig:B_k}. 
This is the opposite of the change identified in~\cite{Kiechle2023}, where $M_\mathrm{eff}$ increased for lower doses up to a certain point and then decreased below the intrinsic value for higher doses.
Since the top layer becomes almost entirely amorphous, and the bottom layer cannot be affected directly by the Ga-ions, we suggest that the effective magnetization change is due to a strain-induced anisotropy~\cite{lin1977contiguous}.

\begin{figure}[tbh!]
    \centering
    \includegraphics[width=0.95\columnwidth]{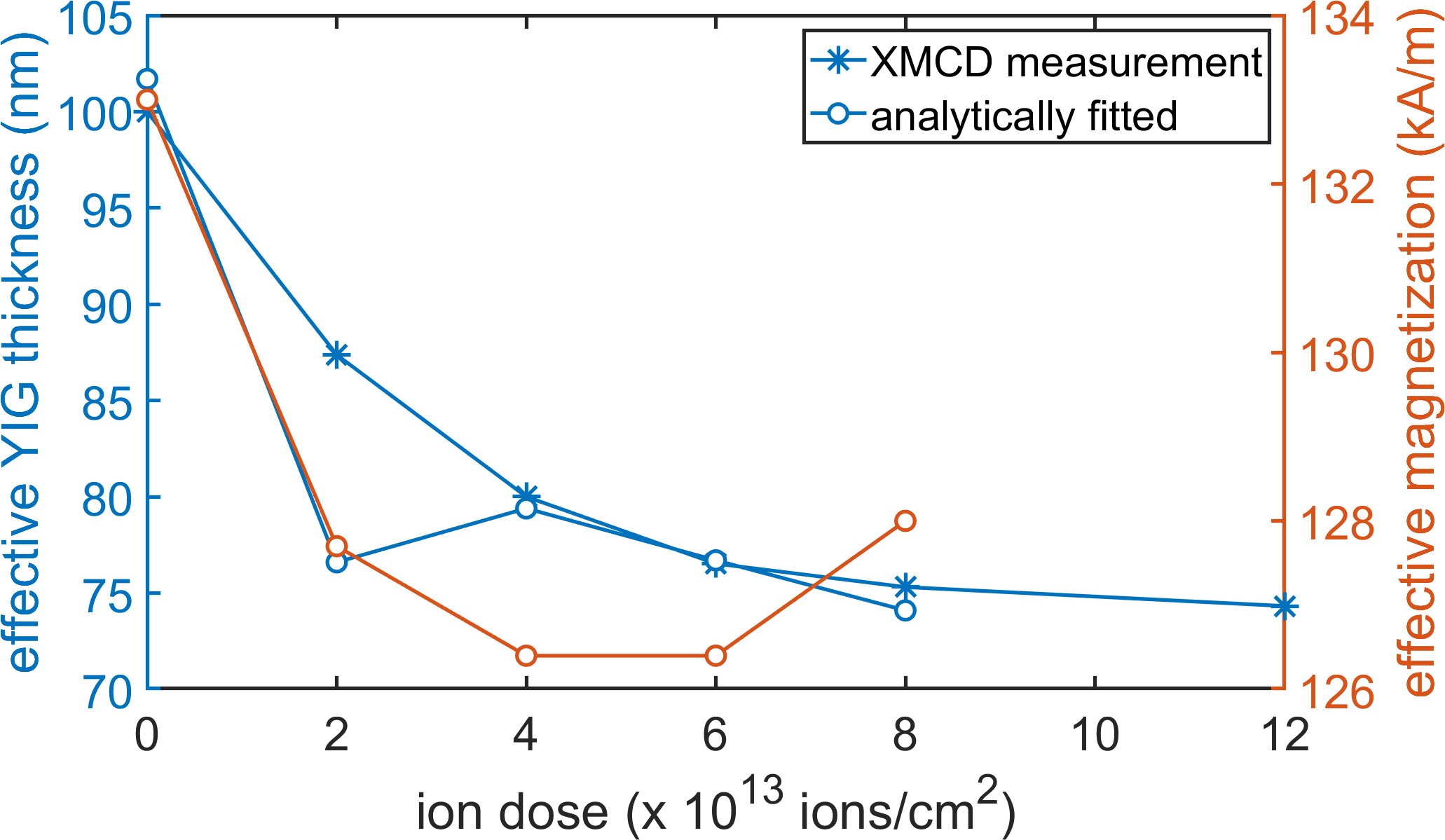}
    \caption{Left ordinate: The change of effective crystalline YIG thickness due to the Ga$^+$-ion irradiation based on the XMCD contrast picture in Fig.\ref{fig:geometry}b). For comparison, the fitted effective YIG thickness based on the analytical dispersion relations shown in Fig.~\ref{fig:B_k} are plotted. The measured effective thickness saturates at around \SI{75}{\nano\meter}, which coincides with the results in~\cite{Kiechle2023} and SRIM simulations. The dose of \SI{12e13}{}\,\ionspercm{} in the alignment field was used in addition to the intended dose map. Right ordinate: Extracted effective magnetization from the fitted dispersion relations in Fig.~\ref{fig:B_k}.}
    \label{fig:TeffAndMeff}
\end{figure}

Using the XMCD contrast pictures of the dose map in Fig.~\ref{fig:geometry}b), it is possible to estimate the change of effective crystalline YIG thickness in the irradiated fields relative to the intrinsic film.
The transmitted light intensity $I$ is calculated via the well-known Lambert-Beer law $ I = I_0\exp(-\sigma t)$, with the incident light intensity $I_0$, and the film thickness $t$.
The material parameter $\sigma = \overline{\sigma} + \sigma_\mathrm{M}$ splits up in a material-related contribution $\overline{\sigma}$ capturing effects like molar or density absorbance that is assumed approximately constant in our experiment, and a magnetic contribution $\sigma_\mathrm{M}$ that is proportional to the magnetization or spin polarization of the material (Fe in this case).
Under the assumption that only $t$ and $\sigma_\mathrm{M}$ may change due to irradiation, the relative change in magnetization and thickness (together: the total magnetic moment) due to irradiation with dose $\mathrm{Dx}$ can be estimated via 
\begin{equation}
\label{eq:Ms_change}
   \frac{M_\mathrm{Dx}}{M_0}\frac{t_\mathrm{Dx}}{t_0} \propto \frac{\ln\left(I_\mathrm{Dx}^+ / I_\mathrm{Dx}^- \right)}{\ln\left( I_0^+/I_0^- \right)},
\end{equation}
where $I_\mathrm{Dx}^\pm$ is the measured light intensity for an external magnetic DC bias field of $\pm\SI{250}{\milli\tesla}$, $t_0$ is the original YIG film thickness, and $t_\mathrm{Dx}$ is the effective magnetic crystalline YIG film thickness in the irradiated area with dose $\mathrm{Dx}$.
Note that from \eqref{eq:Ms_change}, it is not possible to differentiate between a change in magnetization and a change in thickness. 
However, we know from previous studies \cite{Kiechle2023} and SRIM simulations that the Ga$^+$-ion penetration depth is approximately \SI{25}{\nano\meter}, and there is no change in the physical thickness of the film since the ion etching is negligible at these dose levels.
Additionally, the number of implanted Ga$^+$ ions is low compared to the Fe concentration such that the change due to variations, e.g., in the number density, molar absorptivity, or amount concentration represented by $\overline{\sigma}$ is negligible.
Thus, ion irradiation does not directly affect the bottom \SI{75}{\nano\meter} of the YIG film. 
The XMCD contrast change results from the decreasing saturation magnetization of the top irradiated layer. 
The change in total magnetic moment corresponding to an effective thickness of the YIG film that is proportional to the XMCD contrast change is plotted in Fig.~\ref{fig:TeffAndMeff} on the left ordinate.
For the extraction, we additionally use the contrast of the alignment field, which was done at a dose of \SI{12e13}{}\,\ionspercm{}.
The extracted effective thickness shows good agreement with the TEM measurements in~\cite{Kiechle2023} and saturates close to the mean penetration depth of the Ga$^+$-ion at \SI{75}{\nano\meter}.
%The fitted YIG thicknesses in Fig.~\ref{fig:B_k}, which denote the slope of the wavenumber curves, suggest a more abrupt change in effective thickness but agree sufficiently well with the measured values in Fig.\ref{fig:TeffAndMeff}.
The fitted YIG thicknesses in Fig.~\ref{fig:B_k}, which denote the slope of the wavenumber curves, are additionally plotted on the left ordinate in Fig.~\ref{fig:TeffAndMeff}, and agree with the measured values. 
For comparison, we plot the extracted $M_\mathrm{eff}$ from the legend of Fig.~\ref{fig:B_k} on the right ordinate of Fig.~\ref{fig:TeffAndMeff}. 
%The XMCD contrast shows that the total magnetic moment decreases monotonously with increasing Ga$^+$-ion dose as those areas become brighter.
%In this simplified picture, the turning point in extracted wavelength or effective refractive index, as shown in Fig.~\ref{fig:dosemap}, could not be explained.
%However, if we also take into account that the effective crystalline YIG film thickness decreases with increasing irradiation dose, a possible explanation for the turning point is the result of two counteracting effects:
%At low doses, the implanted Ga$^+$ atoms locally induce strain in the YIG crystal as they displace atoms from their lattice places in the crystal.
%In our case, this strain induces an in-plane anisotropy that reduces the local $M_\mathrm{eff}$.
%This effect continues until a dose level where there are enough Ga atoms to damage the crystal sufficiently such that it gets partially amorphous, reducing the effective thickness of the YIG film.
%Note that the total thickness of the YIG film remains unchanged as the top layer of the YIG film gets amorphous but is not removed at this dose levels~\cite{Kiechle2023}.
%The combined effect of decreasing $M_\mathrm{S}$ and $t_\mathrm{Dx}$ is plotted in Fig.~\ref{fig:XMCDandMeff}b) on the right y-axis.
%The relative change in $M_\mathrm{eff}$ in Fig.~\ref{fig:XMCDandMeff} is in line with the $M_\mathrm{eff}$ change that was extracted from the altered wavelengths in Fig.~\ref{fig:dosemap}.
As a result, we observe a trade-off between two effects: At low doses, implanted Ga$^+$-atoms locally induce strain in the YIG crystal as they generate a lattice displacement~\cite{lin1977contiguous}.
This effect continues until a Ga concentration is reached, where the crystal is increasingly damaged and gets amorphous, which means that the effective thickness of the YIG film is reduced, and the strain-induced anisotropy gets partially compensated.
%In an illustrative picture, one can think about these opposing effects as first introducing strain in the crystal (Ga$^+$ implantation) and then releasing this strain (crystal amorphization).

%\comment{Any chance to measure this same sample with the trMOKE for longer wavelengths? Although we would probably not see the boxes excited simultaneously because the refractive index change is just too large for longer wavelengths, maybe we could see them separately and fit Meff that way.}
%\comment{JG: it's not easy to get an adapter for the sample holder. Also, the bond wires of the sample are significantly higher than what we are used to. We'll have to see if there is an option to fit it under the microscope.}

\begin{figure}[tbh!]
    \centering
    \includegraphics[width=0.95\columnwidth]{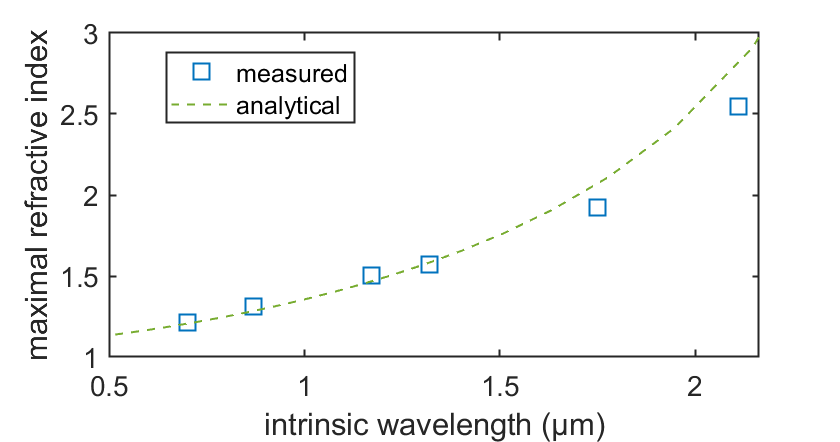}
    \caption{Maximum experimentally achievable refractive index, i.e., the change of wavelength in an irradiated area compared to the intrinsic wavelength, corresponding to the dose of \SI{4e13}{}\,\ionspercm{}.}
    \label{fig:n_max}
\end{figure}

From the perspective of applications, it is instructive to plot the maximum achievable effective refractive index for spin waves against the intrinsic wavelength as shown in Fig.~\ref{fig:n_max}. 
The maximum change in our measurements occurred around \SI{4e13}{}\,\ionspercm{}, so we plot the refractive indices for this dose value. 
Indeed we see the refractive index increase for larger wavelengths at the same dose value and the same $M_\mathrm{eff}$. 
This tendency agrees well with the curve calculated from the analytical dispersion relation, plotted in Fig.~\ref{fig:n_max} as a dashed line. 
The reason for this is that a small change in $M_\mathrm{eff}$ in the dispersion curves corresponds approximately to a shift along the wavenumber axis. 
This shift in $k$ is approximately constant but decreases for larger wavenumbers if expressed as a fraction of $k$. 
More precisely, $n = \frac{k_0}{k}\approx\frac{k_0}{k_0-\Delta{}k}\xrightarrow{ k \to 0 } 1$, where $k_0$ is the intrinsic wavenumber and $\Delta{}k$ is the shift of the dispersion curves in $k$. 

For a rough estimate, we can also state that the required dose for a given refractive index is inversely proportional to the intrinsic wavelength, not considering the nonlinearity of the dispersion relation. 
This can also be deduced from analytical considerations of the dispersion relation.
More importantly, this study demonstrates that one order of magnitude higher dose (compared to \cite{Kiechle2023}) does not hinder the propagation of spin waves. 
The attenuation of the waves in our measurements seems to be negligibly affected, as can be seen in the normalized STXM snapshot of propagating spin waves in Fig.~\ref{fig:snapshot}b). 
This finding allows for further downscaling of spin-wave devices to access the sub-micron wavelength regime. 

\section{Conclusion}
We have measured the effect of Ga$^+$-ion irradiation on short-wavelength spin waves in thin YIG films. 
We observe a non-monotonic change in wavelength, which we attribute to a strain-induced anisotropy in the film in combination with an effective magnetic thickness change. 
We compared our experimental results with analytical calculations using the irradiation modeled as a change in effective magnetization and a decreased effective film thickness due to an amorphized top layer of YIG. 
We observe a decreased effect of irradiation for smaller wavelengths, which the dispersion relation can~directly~explain. 

Our results prove that graded-index FIB patterning can also be used for short-wavelength spin waves, while the required higher doses do not induce substantial losses to the spin wave propagation in the YIG medium. 
The nanoscale patterning of YIG is challenging, although several groups have developed methods to create YIG nanostructures due to their high relevance for magnonics. 
Nevertheless, the FIB irradiation technology's simplicity and ability to create graded-index nanoscale patterns are currently unparalleled among YIG patterning technologies for spin wave devices. 

%\section{Contributions and Acknowledgements}
\section{Acknowledgements}

%J.G. and M.K. performed FIB patterning and pre-characterized the YIG samples. 
%P.N. and J.V. performed e-beam lithography within the framework of this project.
%J.G. and S.W. conducted the STXM measurements. 
%G.C. and M.B. organized and led the project. 
%A.P. and J.G. analyzed the measurement data and wrote the manuscript.
%All authors discussed the results and reviewed the manuscript.

We thank Thomas Meisner, Ulrike Eigenthaler, and Marion Hagel (Max Planck Institute for Intelligent Systems, Stuttgart, Germany) for performing the mechanical thinning, FIB backside etching, and patterning of the transducers, respectively.
%We thank Sebastian Wintz for his valuable support in advance of and during beamtime.
The authors thank Helmholtz-Zentrum Berlin for the allocation of synchrotron radiation beamtime. 
A.P. acknowledges support from the Bolyai Janos Fellowship of the Hungarian Academy of Sciences.

\bibliographystyle{IEEEtran}
\bibliography{IEEEabrv,NiceBib}

\begin{thebibliography}{10}
\providecommand{\url}[1]{#1}
\csname url@samestyle\endcsname
\providecommand{\newblock}{\relax}
\providecommand{\bibinfo}[2]{#2}
\providecommand{\BIBentrySTDinterwordspacing}{\spaceskip=0pt\relax}
\providecommand{\BIBentryALTinterwordstretchfactor}{4}
\providecommand{\BIBentryALTinterwordspacing}{\spaceskip=\fontdimen2\font plus
\BIBentryALTinterwordstretchfactor\fontdimen3\font minus \fontdimen4\font\relax}
\providecommand{\BIBforeignlanguage}[2]{{%
\expandafter\ifx\csname l@#1\endcsname\relax
\typeout{** WARNING: IEEEtran.bst: No hyphenation pattern has been}%
\typeout{** loaded for the language `#1'. Using the pattern for}%
\typeout{** the default language instead.}%
\else
\language=\csname l@#1\endcsname
\fi
#2}}
\providecommand{\BIBdecl}{\relax}
\BIBdecl

\bibitem{Chumak2022}
\BIBentryALTinterwordspacing
A.~V. Chumak, P.~Kabos, M.~Wu \emph{et~al.}, ``Advances in magnetics roadmap on spin-wave computing,'' \emph{IEEE Transactions on Magnetics}, vol.~58, no.~6, p. 1–72, Jun. 2022. [Online]. Available: \url{http://dx.doi.org/10.1109/TMAG.2022.3149664}
\BIBentrySTDinterwordspacing

\bibitem{zenbaa2024}
N.~Zenbaa, C.~Abert, F.~Majcen \emph{et~al.}, ``Magnonic inverse-design processor,'' 2024.

\bibitem{Breitbach2023stimulated}
\BIBentryALTinterwordspacing
D.~Breitbach, M.~Schneider, B.~Heinz \emph{et~al.}, ``Stimulated amplification of propagating spin waves,'' \emph{Physical Review Letters}, vol. 131, no.~15, Oct. 2023. [Online]. Available: \url{http://dx.doi.org/10.1103/PhysRevLett.131.156701}
\BIBentrySTDinterwordspacing

\bibitem{Breitbach2024}
\BIBentryALTinterwordspacing
D.~Breitbach, M.~Bechberger, B.~Heinz \emph{et~al.}, ``Nonlinear erasing of propagating spin-wave pulses in thin-film ga:yig,'' \emph{Applied Physics Letters}, vol. 124, no.~9, Feb. 2024. [Online]. Available: \url{http://dx.doi.org/10.1063/5.0189648}
\BIBentrySTDinterwordspacing

\bibitem{Merbouche_2024True}
\BIBentryALTinterwordspacing
H.~Merbouche, B.~Divinskiy, D.~Gouéré \emph{et~al.}, ``True amplification of spin waves in magnonic nano-waveguides,'' \emph{Nature Communications}, vol.~15, no.~1, February 2024. [Online]. Available: \url{http://dx.doi.org/10.1038/s41467-024-45783-1}
\BIBentrySTDinterwordspacing

\bibitem{wang2024allmagnonic}
Q.~Wang, R.~Verba, K.~Davidkova \emph{et~al.}, ``All-magnonic repeater based on bistability,'' 2024.

\bibitem{wang2024nanoscale}
\BIBentryALTinterwordspacing
Q.~Wang, G.~Csaba, R.~Verba \emph{et~al.}, ``Nanoscale magnonic networks,'' \emph{Physical Review Applied}, vol.~21, p. 040503, Apr 2024. [Online]. Available: \url{https://link.aps.org/doi/10.1103/PhysRevApplied.21.040503}
\BIBentrySTDinterwordspacing

\bibitem{Casulleras2023generation}
\BIBentryALTinterwordspacing
S.~Casulleras, S.~Knauer, Q.~Wang \emph{et~al.}, ``Generation of spin-wave pulses by inverse design,'' \emph{Physical Review Applied}, vol.~19, no.~6, Jun. 2023. [Online]. Available: \url{http://dx.doi.org/10.1103/PhysRevApplied.19.064085}
\BIBentrySTDinterwordspacing

\bibitem{Flebus2024The}
\BIBentryALTinterwordspacing
B.~Flebus, D.~Grundler, B.~Rana \emph{et~al.}, ``The 2024 magnonics roadmap,'' \emph{Journal of Physics: Condensed Matter}, 2024. [Online]. Available: \url{http://iopscience.iop.org/article/10.1088/1361-648X/ad399c}
\BIBentrySTDinterwordspacing

\bibitem{Kiechle2023}
\BIBentryALTinterwordspacing
M.~Kiechle, A.~Papp, S.~Mendisch \emph{et~al.}, ``Spin‐wave optics in yig realized by ion‐beam irradiation,'' \emph{Small}, vol.~19, no.~21, Feb. 2023. [Online]. Available: \url{http://dx.doi.org/10.1002/smll.202207293}
\BIBentrySTDinterwordspacing

\bibitem{Kiechle2022}
\BIBentryALTinterwordspacing
M.~Kiechle, L.~Maucha, V.~Ahrens \emph{et~al.}, ``Experimental demonstration of a spin-wave lens designed with machine learning,'' \emph{IEEE Magnetics Letters}, vol.~13, p. 1–5, 2022. [Online]. Available: \url{http://dx.doi.org/10.1109/LMAG.2022.3209647}
\BIBentrySTDinterwordspacing

\bibitem{Weigand2022}
\BIBentryALTinterwordspacing
M.~Weigand, S.~Wintz, J.~Gr\"{a}fe \emph{et~al.}, ``Timemaxyne: A shot-noise limited, time-resolved pump-and-probe acquisition system capable of 50 ghz frequencies for synchrotron-based x-ray microscopy,'' \emph{Crystals}, vol.~12, no.~8, p. 1029, Jul. 2022. [Online]. Available: \url{http://dx.doi.org/10.3390/cryst12081029}
\BIBentrySTDinterwordspacing

\bibitem{Dubs2020}
\BIBentryALTinterwordspacing
C.~Dubs, O.~Surzhenko, R.~Thomas \emph{et~al.}, ``Low damping and microstructural perfection of sub-40nm-thin yttrium iron garnet films grown by liquid phase epitaxy,'' \emph{Physical Review Materials}, vol.~4, no.~2, Feb. 2020. [Online]. Available: \url{http://dx.doi.org/10.1103/PhysRevMaterials.4.024416}
\BIBentrySTDinterwordspacing

\bibitem{Mayr2021}
\BIBentryALTinterwordspacing
S.~Mayr, S.~Finizio, J.~Reuteler \emph{et~al.}, ``Xenon plasma focused ion beam milling for obtaining soft x-ray transparent samples,'' \emph{Crystals}, vol.~11, no.~5, p. 546, May 2021. [Online]. Available: \url{http://dx.doi.org/10.3390/cryst11050546}
\BIBentrySTDinterwordspacing

\bibitem{Ziegler2010SRIM}
\BIBentryALTinterwordspacing
J.~F. Ziegler, M.~Ziegler, and J.~Biersack, ``Srim – the stopping and range of ions in matter (2010),'' \emph{Nuclear Instruments and Methods in Physics Research Section B: Beam Interactions with Materials and Atoms}, vol. 268, no. 11–12, p. 1818–1823, Jun. 2010. [Online]. Available: \url{http://dx.doi.org/10.1016/j.nimb.2010.02.091}
\BIBentrySTDinterwordspacing

\bibitem{Nolle2012}
\BIBentryALTinterwordspacing
D.~Nolle, M.~Weigand, P.~Audehm \emph{et~al.}, ``Note: Unique characterization possibilities in the ultra high vacuum scanning transmission x-ray microscope (uhv-stxm) “maxymus” using a rotatable permanent magnetic field up to 0.22 t,'' \emph{Review of Scientific Instruments}, vol.~83, no.~4, Apr. 2012. [Online]. Available: \url{http://dx.doi.org/10.1063/1.4707747}
\BIBentrySTDinterwordspacing

\bibitem{Kalinikos1986}
\BIBentryALTinterwordspacing
B.~A. Kalinikos and A.~N. Slavin, ``Theory of dipole-exchange spin wave spectrum for ferromagnetic films with mixed exchange boundary conditions,'' \emph{Journal of Physics C: Solid State Physics}, vol.~19, no.~35, p. 7013–7033, Dec. 1986. [Online]. Available: \url{http://dx.doi.org/10.1088/0022-3719/19/35/014}
\BIBentrySTDinterwordspacing

\bibitem{lin1977contiguous}
Y.~Lin, G.~Almasi, and G.~Keefe, ``Contiguous-disk bubble domain devices,'' \emph{IEEE Transactions on Magnetics}, vol.~13, no.~6, pp. 1744--1764, 1977.

\end{thebibliography}

\end{document}